\documentclass[pdflatex,sn-mathphys-num]{sn-jnl}% Math and Physical Sciences Numbered Reference Style 

\usepackage{graphicx}%
\usepackage{multirow}%
\usepackage{amsmath,amssymb,amsfonts}%
\usepackage{amsthm}%
\usepackage{mathrsfs}%
\usepackage[title]{appendix}%
\usepackage{xcolor}%
\usepackage{textcomp}%
\usepackage{manyfoot}%
\usepackage{booktabs}%
\usepackage{algorithm}%
\usepackage{algorithmicx}%
\usepackage{algpseudocode}%
\usepackage{listings}%
\begin{document}

%% Title (max. 15 words)
\title[Article Title]{\bf The role of spectator modes in the quantum-logic spectroscopy of single trapped molecular ions}

%% Authors 
\author[1]{\fnm{Mikolaj} \sur{Roguski}}
\author[1]{\fnm{Aleksandr} \sur{Shlykov}}
\author[1,2]{\fnm{Ziv} \sur{Meir}}
\author*[1]{\fnm{Stefan} \sur{Willitsch}}\email{stefan.willitsch@unibas.ch}

\affil[1]{\orgdiv{Department of Chemistry}, \orgname{University of Basel}, \orgaddress{\street{Klingelbergstrasse 80}, \city{Basel}, \postcode{4056}, \country{Switzerland}}}
\affil[2]{\orgdiv{\textit{current address}: Department of Physics of Complex Systems}, \orgname{Weizmann Institute of Science}, \orgaddress{\city{Rehovot}, \postcode{761001}, \country{Israel}}}

%% Abstract
%% Abstract (mandatory)
%% ● Should be 150 words or fewer.
%% ● Should be accessible to non-specialists.
%% ● Should include the background and context of the work, then ‘Here we show’ or an equivalent phrase, and then the major
%% results and conclusions of the paper.
%% ● Hyperbolic language (e.g. novel, unprecedented, remarkable, etc.) should be avoided.
%% ● Acronyms should be avoided wherever possible.
%% ● Must not contain citations to references.
%% ● Any key species, protein names, and gene names should be mentioned in the title and/or abstract to ensure optimal retrieval of the paper in database searches.

\abstract{Quantum-logic spectroscopy has become an increasingly important tool for the state detection and readout of trapped atomic and molecular ions which do not possess easily accessible closed-cycling optical transitions. In this approach, the internal state of the target ion is mapped onto a co-trapped auxiliary ion. This mapping is typically mediated by normal modes of motion of the two-ion Coulomb crystal in the trap. The present study investigates the role of spectator modes not directly involved in a measurement protocol relying on a state-dependent optical-dipole force. We identify a Debye-Waller-type effect that modifies the response of the two-ion string to the force. We show that cooling the spectator modes of the string allows for the detection of the rovibrational ground state of an N$_2^+$ molecular ion with a computed statistical fidelity exceeding 99.99\%, improving on previous experiments by more than an order of magnitude while also halving the experimental time. This enhanced sensitivity enables the simultaneous identification of multiple rotational states with markedly weaker signals.}

%% Keywords 

%% Print Title
\maketitle

%% INTRODUCTION 
%% Introduction (mandatory):
%% ● Must begin with the heading ‘Introduction’.
%% ● Should include the background and rationale for the work.
%% ● The final paragraph should be a brief summary of the major results and conclusions.

\section{Introduction}\label{sec:introduction}

Quantum-logic spectroscopy (QLS) protocols provide a framework for the readout of the internal state of a target (‘spectroscopy’) ion by mapping it onto the state of a co-trapped axillary (‘logic’) ion \cite{schmidt05a}. This technique is usually employed for systems without readily accessible optical cycling transitions, e.g., atomic species such as Al$^+$ \cite{schmidt05a}, highly-charged ions \cite{micke20a} and molecular ions \cite{wolf16a,chou17a,sinhal20a,holzapfel24a}. The state mapping is typically mediated by a motional degree of freedom common to the ions which are strongly coupled by the Coulomb interaction. In general, a QLS experimental sequence consists of three steps: first, cooling of a common (target) motional mode, second, manipulation of the target mode population depending on the internal state of the spectroscopy ion and third, motional-state readout by spectroscopy on the logic ion. 

The QLS implementations demonstrated so far used different approaches to project the state of the spectroscopy ion onto the motional state. Several experiments involved adding a phonon to a target mode initially cooled to its ground state by driving a transition on a motional sideband of a spectroscopic transition in the spectroscopy ion \cite{hume07a, holzapfel24a, chou17a, micke20a}. In such a scheme, the addition of a phonon to the target mode indicated a successful state detection. In another approach, the target mode was prepared in the first excited motional Fock state, $|1\rangle$, and depending on the internal state of the spectroscopy ion, a single phonon was exchanged with another mode \cite{wolf16a}. This mode transfer was mediated by a state-dependent optical dipole force (ODF) resonantly driving the mode exchange. Both of these approaches involved single-phonon excitations. A third method relies on a modulated state-dependent ODF tuned to resonance with the target motional mode, thus causing motional excitation \cite{hume11a, sinhal20a}. Depending on the duration of the ODF pulse and the state of the spectroscopy ion, the target mode can thus be excited to highly excited motional states. Various alternative implementations of QLS were proposed \cite{mur-petit12a, loh14a, hudson18a, campbell20a}.  

Here, we further explore the third QLS method mentioned above, which employs a state-dependent ODF to determine the internal state of the spectroscopy ion \cite{hume11a,meir19a,sinhal20a,najafian20b}. The ODF is modulated at a frequency resonant with a normal mode of the two-ion system (referred to as target mode) and its amplitude depends on the internal state of the spectroscopy ion. The resulting resonant motional excitation of the system's motional mode is probed by the logic ion using sideband spectroscopy \cite{meekhof96a, leibfried03a} and thus provides information about the state of the spectroscopy ion.

Cooling the target mode close to its quantum-mechanical ground state is a prerequisite for most implementations of QLS mediated with motional states. However, population in the remaining normal modes which do not directly participate in the measurement protocol (spectator modes) could have detrimental effects on the spectroscopy due to Kerr-like \cite{roos08a}, mode-cross-talk \cite{wineland98a, marquet03a} or Debye-Waller (DW) \cite{wineland98a} effects. Consequently, it is common to set the frequencies of all the modes to be well-separated (i.e. avoid any resonances), and to cool all modes to sub-Doppler temperatures \cite{chou17a, marshall25a}. Here, we study in detail the consequences for QLS of the DW effect that modifies light-ion interaction strength due to motion of the ion. This effect was shown to be of relevance for applications in precision spectroscopy, quantum simulations, and quantum computing with multiple ions \cite{wineland98a, schmidt05a, bermudez17a, chen17a, bruzewicz19a, affolter20a}. 

DW effects are present in two stages of the present QLS method. First, population in the spectator modes may alter the interaction of the molecule with the ODF pulse. Second, it also affects the readout of the motional excitation with sideband spectroscopy, leading to a reduced signal-to-noise ratio and, thus, decreased state-detection fidelities. While the second effect was shown to be small in previous studies \cite{sinhal20a}, the first can have significant ramifications for the state-detection fidelity, as shown here.
 
We experimentally observe differences in response to the ODF pulse when the two-ion Coulomb crystal was prepared in specified spectator mode states. We rationalized these differences with the aid of analytical models and numerical simulations. We show that the spectator modes play an important role during the motional excitation via the DW effect. By cooling the axial spectator mode close to its ground state, we achieve an improved computed statistical state-detection fidelity, exceeding 99.99\% for as few as nine experimental repetitions. Compared to 99.5\% after 22 cycles reported previously when only the target mode was cooled \cite{sinhal20a}, the execution time of the protocol was halved, while simultaneously achieving a more precise state determination. Furthermore, we show that the improved sensitivity of the detection method can also assist in the identification of higher-lying rotational states of the molecule. 

%% EXPERIMENTAL METHODS 
\section{Methods}\label{sec:exp_methods}

\begin{figure}[ht] 
    \centering 
    \includegraphics[width=0.90\textwidth]{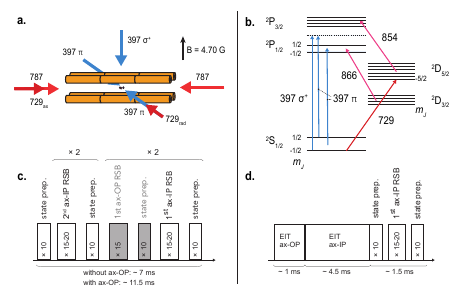} 
    \caption{\textbf{Experimental scheme.} \textbf{a.} Schematic of the ion trap with relevant laser beams indicated. \textbf{b.} Partial Zeeman-resolved energy level scheme of $^{40}$Ca$^+$ with spectroscopic transitions used for EIT cooling (blue) and sideband cooling (red) indicated, together with two repumping lasers (pink). \textbf{c.} Typical pulsed sideband cooling sequence. Each box represents repeated pulses on a red sideband of a $(4s)^2S_{1/2}(m_j=-1/2) \rightarrow (3d)^2 D_{5/2}(m_j=-5/2)$ transition in Ca$^+$  or pulses on a $(4s)^2S_{1/2}(m_j=+1/2) \rightarrow (3d)^2D_{5/2}(m_j=-3/2)$ transition to prepare Ca$^+$ in a $S_{1/2}(m_j=-1/2)$ state. The pulses were followed by repumping with a laser beam at 854 nm. \textbf{d.} Typical multi-stage EIT cooling sequence. See text for details.}
   \label{fig:ions_energy_lev}
\end{figure}

The experimental sequence for the quantum-state-detection of N$_2^+$ consisted of four stages: state preparation, translational cooling, motional excitation and detection of the resultant motional state. The experimental details were described previously in Refs. \cite{meir19a} and \cite{sinhal20a}. Here, only the main points are recapitulated. 

Single $^{14}$N$_2^+$ ions were prepared in the rovibrational ground state using a 2+1$'$ resonance enhanced multi-photon ionization (REMPI) scheme of internally cold neutral N$_2$ molecules from a supersonic molecular beam \cite{shlykov23a}.
The molecular ions were loaded into a linear trap operated at a radiofrequency of 19.1~MHz and endcap voltages of $\sim$ 150 V. The motional frequencies for the normal modes of the N$_2^+$-Ca$^+$ ion string, $\omega_m$, were measured using spectroscopy of motional sidebands on the Ca$^+$ $(4s)^2S_{1/2}(m_j=-1/2) \rightarrow (3d)^2D_{5/2}(m_j=-5/2)$ Zeeman-resolved `clock' transition around 729~nm and are presented in Table \ref{tab:frequencies_and_temperatures}. A static external magnetic field of 4.70~G was applied to define the quantization axis.

The motion of a two-ion crystal in a linear radiofrequency ion trap (see schematic in Fig. \ref{fig:ions_energy_lev}) can be described in terms of six normal modes. Two modes, with in-phase (ax-IP) and out-of-phase (ax-OP) motions of the ions, are directed along the longitudinal trap axis. The other four modes are directed along two perpendicular radial principal axes of the trap (rad-OP and rad-IP along each axis). Due to small asymmetries of the radial potential, the corresponding radial mode frequencies usually differ slightly ($\sim$ 10 kHz here).

\begin{table}[htbp]
\caption{Mode frequencies, $\omega_m$, and minimum mean phonon numbers, $\bar{n}$, after cooling of the motional modes of the N$_2^+$-Ca$^+$ Coulomb crystal. The relevant cooling methods are indicated in the last column. The radial modes along different axes are denoted with $>$($<$). If not specified, `SB$_{\text{no}\, \text{ax-OP}}$' refers to sideband cooling without addressing the ax-OP mode, `SB$_{\text{1}\times \text{ax-OP}}$' (`SB$_{\text{2}\times \text{ax-OP}}$') to sideband cooling with one (two) sets of ax-OP sideband pulses, and `EIT$_{2}$' (`EIT$_{3}$') multi-stage electromagnetically induced transparency cooling of both the in- and out-of-phase modes along two (all three) principal axes of the trap. See text for details.} 
\label{tab:frequencies_and_temperatures}
\begin{tabular}{@{}llll@{}}
\toprule
Mode & $\mathbf{\omega}_m$   & $\bar{n}$ & Method\\
\midrule
ax-IP         & 674 kHz            & 0.14(3)                       & SB, EIT$_{2}$ and EIT$_{3}$ \\      
ax-OP         & 1204 kHz              & $\sim$ 8                      & SB$_{\text{no}\, \text{ax-OP}}$    \\
              &                       & $\sim$ 3                      & SB$_{\text{1}\times \text{ax-OP}}$ \\
              &                       & $<$ 1                         & SB$_{\text{2}\times \text{ax-OP}}$ \\
              &                       & $<$ 0.5                       & EIT$_{2}$ and EIT$_{3}$     \\ 
rad-OP$_{<}$  & 545 kHz             & 0.6(2)                        & EIT$_{2}$ and EIT$_{3}$     \\ 
rad-OP$_{>}$  & 556 kHz             & $<$ 1                         & EIT$_{3}$            \\
              &                       & $>$ 7                         & EIT$_{2}$            \\ 
rad-IP$_<$        & 1040 kHz       & $<$ 1         & EIT$_2$ and EIT$_{3}$            \\
rad-IP$_>$      & 1051 kHz         & $<$ 1          &    EIT$_3$ \\
\botrule
\end{tabular}
\end{table}

\subsection{Cooling methods}\label{sec:cooling_methods}
Schematics of the geometric configuration of the lasers with respect to the ion trap as well as the energy levels and transitions used for the present cooling schemes are depicted in Figs. \ref{fig:ions_energy_lev} (a) and (b), respectively. Laser cooling was performed on Ca$^+$ in two stages. First, all six modes of the two-ion crystal were cooled close to the Doppler limit by driving a red-detuned $(4s)^2S_{1/2} \rightarrow (4p)^2P_{1/2}$ transition with a 397~nm laser beam with a 45$^{\circ}$ projection along the axial and 60$^{\circ}$ projection to the radial axes. Another laser at 866~nm served to repump population from the metastable $(3d)^2D_{3/2}$ level, closing the cooling cycle. Next, the ions were prepared close to the motional ground state of the target (ax-IP) mode using different secondary cooling schemes: 
\begin{itemize}
\item
\textit{Pulsed sideband} cooling (PSC) utilises a train of excitation pulses on the second and first-order sidebands on a narrow transition \cite{leibfried03a}, here ax-IP red sidebands (RSB) on the Ca$^+$ clock transition. A repumper laser pulse at 854~nm followed each of the sideband pulses to pump the ion from the metastable $(3d)^2D_{5/2}$ state to the $(4p)^2P_{3/2}$ state, from where it decayed back to the electronic ground state (Fig. \ref{fig:ions_energy_lev} (b)). The PSC sequence is shown in Fig. \ref{fig:ions_energy_lev} (c). The number and duration of each pulse were empirically optimised to provide the best cooling performance. Moreover, if required, either one (SB$_{\text{1}\times \text{ax-OP}}$) or two (SB$_{\text{2}\times \text{ax-OP}}$) additional cooling pulse sequences on the ax-OP (spectator) mode were interlaced in the SB cooling sequence to reduce the population in this mode down to three or below one phonon. If no pulses were added (SB$_{\text{no}\, \text{ax-OP}}$), the average ax-OP mode population was close to the Doppler limit. 
\item
\textit{EIT cooling} \cite{morigi00a, roos00a, lechner16a} was realized on Zeeman components of the 397~nm transition between the $(4s)^2S_{1/2}$ and $(4p)^2P_{1/2}$ levels with a $\sigma^{+}$-polarized coupling beam and two counter-propagating $\pi$-polarized cooling beams (Fig. \ref{fig:ions_energy_lev} (b)). Each of the cooling laser beams provided a strong projection of its $k$-vector along one of the radial directions and at 60$^{\circ}$ to the axial direction. In this configuration, the axial modes were always cooled. However, by using only one or both EIT $\pi$-polarized cooling beams, modes along either one (EIT$_{2}$) or both radial axes (EIT$_{3}$) were efficiently addressed.
\end{itemize}

In our experiments, the motional modes (see Table \ref{tab:frequencies_and_temperatures}) could be divided into two frequency groups -- first, around 600 kHz, with ax-IP and rad-OP modes, and second, around twice the frequency of the first group with ax-OP and rad-IP. Typically, two EIT pulses were implemented to cool each of the groups efficiently \cite{lechner16a, scharnhorst18a}. In Fig.~\ref{fig:ions_energy_lev}~(d), an example of such a multi-stage EIT cooling sequence is presented. Each pulse is annotated with the axial mode targeted for optimal cooling.

It was experimentally confirmed that, given sufficient EIT cooling time, axial and rad-OP modes could be efficiently cooled below one phonon. The EIT parameters for each stage were adjusted empirically to provide optimal cooling for the modes of interest. EIT cooling reduced the population of the ax-IP mode to $\bar{n} \sim$ 0.4 phonons. To cool this mode even further, the EIT sequence was followed by a set of first-RSB pulses similar to the ones used in PSC to ensure a similar minimum temperature of the ax-IP mode across all cooling methods.

\subsection{Motional mode populations after cooling}

The performance of the cooling methods for different modes of the two-ion crystal was assessed by measuring the final mode populations using the SB-ratio method \cite{diedrich89a, turchette00a}, which yields satisfactory estimates for thermal mode populations close to the ground state. The results are presented in Table \ref{tab:frequencies_and_temperatures}. The ax-OP state population after SB$_{\text{no}\, \text{ax-OP}}$, was calculated from the Doppler-limit temperature and was in agreement with the estimates obtained with the SB-ratio method \cite{metcalf99a, leibfried03a}.

The heating rates in our trap were measured to be $\sim$ 0.05 phonons/ms for the ax-IP and $\sim$ 0.1 phonons/ms for the ax-OP mode \cite{home13a}. These heating rates are negligible compared to the timescales of the present experiments. The temperature and heating rates of the radial modes were observed to have a negligible effect on the axial excitation dynamics due to their negligible Lamb-Dicke factors with the lattice lasers which were generally directed along the longitudinal trap axis.

Generally, the 729 nm interrogation laser beam was aligned along the longitudinal trap axis thus only addressing the axial modes. For diagnostics of the radial modes, we aligned part of the laser beam to a 60$^\circ$ projection onto the radial modes (see Fig.~\ref{fig:ions_energy_lev} (a)). The average phonon number after cooling $\bar{n}$ for the rad-IP modes quoted in Table~\ref{tab:frequencies_and_temperatures} is an estimate based on the expected performance of the EIT cooling for this mode.

\subsection{Motional excitation}

The ODF generating the motional excitation of the two-ion string was realized using a travelling lattice generated by two counter-propagating $\pi$-polarized laser beams with a frequency difference of the IP axial mode between them. The wavelength of the lattice laser (787.4505~nm referenced to a HighFinesse WSU-30 wave meter) was detuned by 12~GHz from the $R_{11}(1/2)$ $ A^2\Pi_u(v'=2) \leftarrow X^2\Sigma_g^+(v''=0)$ rovibronic transition in N$_2^+$. Because the ac-Stark effect causing the ODF is inversely proportional to the detuning from the transition, the lattice interacts strongest with, and is thus most sensitive to, a molecule in the ground rovibronic state \cite{sinhal20a, najafian20a}.
Before applying the lattice lasers, the Ca$^+$ ion was shelved in the $(3d)^2D_{5/2}(m_j=-5/2)$ level to suppress spurious ODF generated by the atomic ion coupling to the $(4p)^2P$ states \cite{sinhal20a}. The ODF pulse length was set to 500 $\mu$s for all experiments presented here, except of the example in Fig. \ref{fig:figure2} (a) for which it was 1 ms and the experiments in Fig. \ref{fig:higher_rotational_states} for which it was 2 ms. 

\subsection{State detection}

\begin{figure}[ht] 
    \centering \includegraphics[width=1.0\textwidth]{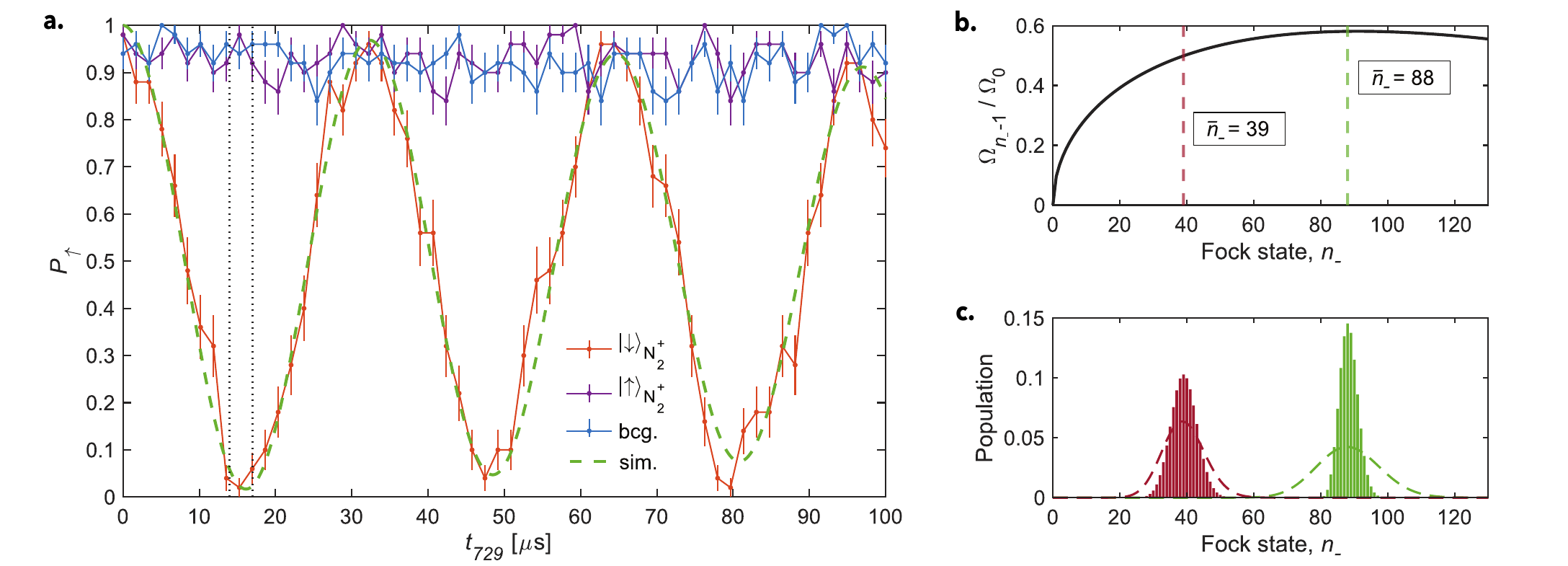} 
    \caption{\textbf{Molecular state detection.} {\bf a.} Rabi oscillations on the blue sideband of a narrow $D_{5/2}(m_j=-5/2) \rightarrow S_{1/2}(m_j=-1/2)$ transition in Ca$^+$ after applying an ODF pulse of 1000~$\mu$s duration when N$_2^+$ was in the ground rotational state (orange trace) and in an excited state (violet trace). The population of the spectator mode was $\bar{n}_+\lesssim$ 0.5 phonon. The background signal (blue trace) was obtained without applying an ODF. Uncertainties represent the standard error of the mean. The black dashed lines indicate the interval of Rabi times in which the signal-to-noise ratio is highest for the molecular-state determination. The green dashed line represents a simulation of the Rabi flop. {\bf b.} Rabi frequencies of the flops on the blue sideband shown in {\bf a.} as a function of Fock-state quantum number $n_-$. The red and green dashed lines indicate the average Fock state of the motional wavepackets shown in {\bf c.} generated by motional excitation through the ODF under different experimental conditions. See text for details.} 
    \label{fig:figure2}
\end{figure}

The state-dependent ODF projected the internal molecular state on the ax-IP motional state of the two-ion crystal, which was subsequently probed by Rabi blue-sideband spectroscopy on the Ca$^+$ clock transition \cite{sinhal20a}. An example blue-sideband Rabi flop is shown in Fig. \ref{fig:figure2} (a). When the ax-IP mode is initially cooled to its motional ground state, a Rabi flop is observed only when motional excitation occurs, i.e., when the molecule is in the rovibronic ground state (orange trace). When N$_2^+$ is in an excited state, there are no oscillations (violet trace), similar to the background signal (blue).

As the probe laser was aligned along the trap axis, only the axial modes contribute directly to the signal. The target (ax-IP) mode population, $n_-$, and the spectator (ax-OP) mode population, $n_+$, affect the Rabi frequency for a blue-sideband transition in a two-level system \cite{wineland98a, meir19a}: 
\begin{equation}
    \label{eq:rabi_freq_thermometry}
    \Omega_{n_+, n_- - 1} = \Omega_0 e^{-\eta_+^2 / 2} L_{n_+}^0 (\eta_+^2) e^{-\eta_{-}^2 / 2} n_-^{-1/2} \eta_- L_{n_--1}^{1} (\eta_{-}^2) , 
\end{equation}
where $\Omega_0$ is the bare Rabi frequency and $L^{\alpha}_{n}$ a generalised Lauguerre polynomial. Note that since the Ca$^+$ is initialized in the excited $D_{5/2}$ electronic state, the blue sideband corresponds to a change in the target mode of $n_{-}-1$.
The Lamb-Dicke parameter for the target mode ($-$) is $\eta_{-} = k z_{-}^{0} \cos(\theta)$, and for the spectator mode ($+$) is $\eta_{+} = k z_{+}^{0} \sin(\theta)$, taking values 0.096 and 0.051 respectively. The spatial extent of the motional-ground-state wavefunction for Ca$^+$ for each mode is
\begin{equation}
    \label{eq:spatial_extent_wf}
    z_{\pm}^{0}=\sqrt{\hbar/(2 m_2 \omega_{\pm})} \,,
\end{equation}
where the frequencies of the two normal modes are given by \cite{morigi01a, wuebbena12a}:
\begin{equation}
    \label{eq:mode_freq_eq1}
    \omega_{\pm}^2=\omega_2^2\left(1+\mu \pm \sqrt{1+\mu^2-\mu}\right),
\end{equation}
with $\omega_2$ the axial frequency of a single Ca$^+$ ion of mass $m_2$ and the mass ratio of the ions is given by $\mu = \frac{m_2}{m_1}$. The mass-dependent factor $\theta$ is defined as:
\begin{equation}
    \label{eq:tan_theta_eq}
    \tan (\theta)=1 / \sqrt{\mu}-\sqrt{\mu}+\sqrt{1 / \mu+\mu-1} .
\end{equation}
In Eq. \ref{eq:rabi_freq_thermometry}, the populations in the spectator modes introduce corrections to the Rabi frequency in the form of a DW factor \cite{wineland98a}:
\begin{equation}
\label{eq:DW_plus}
DW_{+} = e^{-\eta_+^2 / 2} L_{n_+}^0 (\eta_+^2) .
\end{equation}

The excited-state population $P_{\uparrow}$ for a two-level system undergoing Rabi oscillations is weighted over the probabilities of the ion occupying different motional states, $P_{n_+, n_-}$, and can be written as \cite{wineland98a}: 
\begin{eqnarray}
    \label{eq:rabi_oscillations_eq1}
    P_{\uparrow}(\delta, t) = \sum_{n_+, n_-} P_{n_+, n_-} 
    \frac{\Omega_{n_+, n_{-}-1}^2}{\Omega_{n_+, n_- - 1}^2 + \delta^2}
    \sin^2 \left( \sqrt{\Omega_{n_+, n_{-} - 1}^2 + \delta^2} \, t / 2 \right) ,
\end{eqnarray}
where $t$ is the pulse time and $\delta$ is the detuning of the excitation laser frequency from the blue-sideband transition (in the present experiments, $\delta=0$). The $P_{n_+, n_-}$ probabilities depend on the applied cooling methods, on heating effects, and on the characteristics of the motional excitation, as elaborated in Sec. \ref{sec:simulations}.

Additionally, the Rabi flop is subject to decoherence. Unlike the lifetime of the excited metastable $^2D_{5/2}$ state of Ca$^+$ that is of the order of a second, the decoherence time for the motional sideband transition when trapping a two-ion string, $T_{2}$, is around 500 $\mu$s and has to be included in the fitting of experimental Rabi flops. To account for this, and for the fact that in the presented experiments Ca$^+$ was initially shelved in $^2D_{5/2}(m_j=-5/2)$, the excited-state probability used for fitting the experimental data is \cite{wineland98a}:
\begin{equation}
    \label{eq:probability_excited_state_decoherence}
    P_{\uparrow}^{dec.}(\delta, t, T_2)=1-\big[(P_{\uparrow}(\delta, t)e^{-t/T_{2}}+(1-e^{-t/T_{2}})/2)\big].
\end{equation}

%% Simulations 
\section{Simulations}\label{sec:simulations}

To interpret the experimental results, we performed both classical and quantum simulations of the two-ion motional-excitation dynamics when applying the ODF with the travelling lattice. The simulations yield motional population distributions, represented as $P_{n_+, n_-}$ from Eq. (\ref{eq:rabi_oscillations_eq1}), for different lattice parameters. Here, we briefly discuss the theory underlying the simulations. Although solutions for two ions are presented for generality, we emphasize that in the experiments and simulations, only N\(_2^+\) interacted with the lattice.

The confining potential was modelled as harmonic. Trap anharmonicities may become relevant for high motional states, leading to various types of mode-coupling and/or shifts in the motional frequencies \cite{home11a}. However, as shown experimentally, such effects are considered to be small for the present trap and beyond the current level of experimental precision.

Since the radial modes are perpendicular to the axial direction of propagation of the travelling lattice, they are not expected to contribute to the excitation dynamics (as confirmed experimentally, see Fig. \ref{fig:rabi_flops_different_cooling} (b)). Hence, the simulations are restricted to the axial direction.

\paragraph*{Classical Simulations:}
The classical one-dimensional equation of motion for each ion $j=\{1,2\}$ of mass $m_j$ interacting with the travelling-optical-lattice pulse is given by \cite{meir19a}:
\begin{equation}
    \label{eq:classical_eom}
    \ddot{z}_j = -\omega_j^2z_j \mp \frac{1}{m_j}\frac{e^2/4\pi\varepsilon_0}{(z_2-z_1)^2}+\frac{4k}{m_j}\Delta E_{ac}^{0,j} \sin(2kz_j-\omega_l t) ,
\end{equation}
where $\omega_j$ is the frequency of ion $j$ in the trap, $k$ is the wavenumber of a single lattice laser, $\omega_l$ is the frequency difference between the lattice lasers, $e$ is the elementary charge, $\varepsilon_0$ is the vacuum permittivity, and $z_j$ are the $z$-coordinates of the ions. $\Delta E_{ac}^{0,j}$ is the ac-Stark shift on ion $j$ exerted by a single lattice-laser beam. The terms in the equation correspond to the interaction of the ion with the trapping potential, with the co-trapped ion, and with the optical lattice, respectively. The sign before the second term is `-' for $j$=1 and `+' for $j$=2 respectively.

Eq. (\ref{eq:classical_eom}) was solved using a 4\textsuperscript{th} order Runge-Kutta algorithm \cite{runge1895a} to calculate the kinetic energy of the ions after time duration of $t=t_E$ . Initially ($t=0$), the ions are considered to be in their equilibrium positions with zero velocity. 

\paragraph*{Quantum Simulations:} 

To describe the one-dimensional quantum dynamics of the system, it is helpful to express the motion of the two-ion crystal in the trap in terms of the two axial normal modes. The time-dependent Hamiltonian of the two ions interacting with the optical lattice is given by \cite{najafian20b}:
\begin{eqnarray}
    \label{eq:quantum_hamiltonian1}
        \hat{H}=  &&\hbar\omega_{-}(\hat{a}^\dagger_{-}\hat{a}_{-} + \frac{1}{2}) +\hbar\omega_{+}(\hat{a}^\dagger_{+}\hat{a}_{+} + \frac{1}{2})\nonumber\\
        &&+ \sum_{j=1,2} 2\Delta E_{ac}^{0,j}(1+\cos(2k\hat{z}_j-\omega_l t)) ,
\end{eqnarray}
where $\hat{a}_{m}(^\dagger)$ is the creation (annihilation) operator of mode $m$ and $\hat{z}_j$ is the position operator of ion $j$. 

The time-evolution of the system follows the Schr\"{o}dinger equation, $i\partial_t\psi(t) = \hat{H}\psi(t)$. The motional states can be described in terms of the wave function in the interaction picture, $\psi_I(t)=\sum_{n_+,n_-}C_{n_+,n_-}(t) |n_+, n_-\rangle$, with the time-dependent coefficients $C_{n_+,n_-}(t)$ defined on the two-dimensional Fock space of both axial eigenmodes. 

Assuming only single-phonon transitions in the target mode, we derive an analytical expression for the time evolution of the coefficients (see Appendix \ref{sec:appendix1} for details):
\begin{align}
\label{eq:quantum_analytical_sol}
i\hbar\dot{C}_{n_+,n_-}=&C_{n_+,n_--1}e^{-i\delta_- t}\frac{1}{\sqrt{n_-}}\Big(\Delta E_{ac}^{0,1} \, e^{+i\phi_1} DW_{+}^{(1)} e^{-(\eta_{-}^{(1)})^2/2}\eta_{-}^{(1)} L_{n _--1}^1((\eta_{-}^{(1)})^2) + \nonumber\\ 
&\Delta E_{ac}^{0,2} \, e^{+i\phi_2} DW_{+}^{(2)} e^{-(\eta_{-}^{(2)})^2/2} \eta_{-}^{(2)}L_{n_- -1}^1((\eta_{-}^{(2)})^2) \Big) \nonumber\\
&+ C_{n_+,n_- +1}e^{+i\delta_- t}\frac{1}{\sqrt{n_- +1}}\Big(\Delta E_{ac}^{0,1} \, e^{-i\phi_1} DW_{+}^{(1)} e^{-(\eta_{-}^{(1)})^2/2} \eta_{-}^{(1)} L_{n_-}^1((\eta_{-}^{(1)})^2) \nonumber\\ 
&+  \Delta E_{ac}^{0,2} \, e^{-i\phi_2} DW_{+}^{(2)} e^{-(\eta_{-}^{(2)})^2/2} \eta_{-}^{(2)} L_{n_-}^1((\eta_{-}^{(2)})^2) \Big) ,
\end{align}
where $\delta_{-}$ is the detuning between the modulation frequency of the lattice and the frequency of the target mode, and $\phi_j$ is the phase shift of the lattice on ion $j$. The Lamb-Dicke parameters defined with respect to both modes for each ion are given by:
\begin{eqnarray}
& \eta_{+}^{(1)}=2 k z_+^{0} \sqrt{\mu} \cos (\theta) , \label{eq:new_def_LD_params1}\\
& \eta_{-}^{(1)}=2 k z_-^{0} \sqrt{\mu} \sin (\theta) ,\label{eq:new_def_LD_params2}\\
& \eta_{+}^{(2)}=-2 k z_+^{0} \sin (\theta) , \label{eq:new_def_LD_params3}\\
& \eta_{-}^{(2)}= 2 k z_-^{0} \cos (\theta) . \label{eq:new_def_LD_params4}
\end{eqnarray}
Note the factor of two which originates from using two counter-propagating lasers forming the travelling lattice. The mass-dependent factor $\theta$ is defined according to Eq.~(\ref{eq:tan_theta_eq}). The Lamb-Dicke parameters thus defined for the presented experiments took values $\big(\eta_{+}^{(1)}, \eta_{-}^{(1)}, \eta_{+}^{(2)}, \eta_{-}^{(2)}\big)$ = (0.160, 0.150, -0.094, 0.179) for the presented experiments.

The DW factors of the spectator mode in Eq. (\ref{eq:quantum_analytical_sol}) are defined as:
\begin{equation}
DW_{+}^{(1)} = e^{-(\eta_{+}^{(1)})^2 / 2} L_{n_+}^0 ((\eta_+^{(1)})^2) , \label{eq:DW1}
\end{equation}
\begin{equation}
DW_{+}^{(2)} = e^{-(\eta_{+}^{(2)})^2 / 2} L_{n_+}^0 ((\eta_+^{(2)})^2) . \label{eq:DW2}
\end{equation}
From this definition it becomes apparent that the DW factors act as damping factors for the motional excitation which depend on the spectator-mode populations. In Sec.~\ref{sec:results}, we will see their significant effect on the motional Fock-state distribution after applying the ODF.

To obtain $C_{n_+,n_-}$ and hence calculate the mode populations, $P_{n_+, n_-} = |C_{n_+,n_-}|^2$, the Schrödinger equation can be solved either directly numerically starting from the Hamiltonian Eq. (\ref{eq:quantum_hamiltonian1}) and using, e.g., the Qutip package \cite{johansson12qutip}, or by solving the system of differential equations Eq. (\ref{eq:quantum_analytical_sol}). The latter method is significantly less computationally expensive. However, it assumes only single-phonon transitions in the target mode, no spectator-mode excitation during the dynamics and no cross-talk between the modes. These conditions are satisfied when the detuning of the optical lattice from the target-mode frequency is smaller than the detuning from the spectator mode and both normal mode frequencies, i.e. $\left|\delta_-\right| \ll\left|\delta_+\right|,\,\,\omega_{-},\,\omega_{+}$. 

These assumptions may break down for high average phonon populations and broader phonon distributions when the probability of exciting non-linear resonances may become significant \cite{wineland98b}. Hence, we tested the equivalence of the two simulation methods for a relevant range of excitation times and ac-Stark shifts and observed no significant discrepancy between the models. Hence, we used the faster method capitalizing on Eq. (\ref{eq:quantum_analytical_sol}) to obtain the motional state distributions required for simulating the experimental blue-sideband Rabi oscillations Eq. (\ref{eq:probability_excited_state_decoherence}). In the simulations, the ions were initially in the ground state of the target mode with different populations on the OP spectator mode.

In the considered model, we do not include non-linear effects originating from the higher-order terms in the Coulomb interaction between the ions \cite{roos08a, nie09a}. Such ``Kerr-like" effects are known to modulate ax-OP frequencies by coupling to rad-OP modes (or rad-OP modes to each other) in the case of equal-mass two-ion strings \cite{ding17a}. Also, if these modes fall into parametric resonance conditions, such as $f_{ax-OP} \equiv 2 f_{rad-OP}$, the interaction anharmonicities can lead to a non-negligible population transfer.

\section{Results and discussion}\label{sec:results}

\begin{figure}[htbp]
    \centering % Center the figure
    \includegraphics[width=1.0\textwidth]{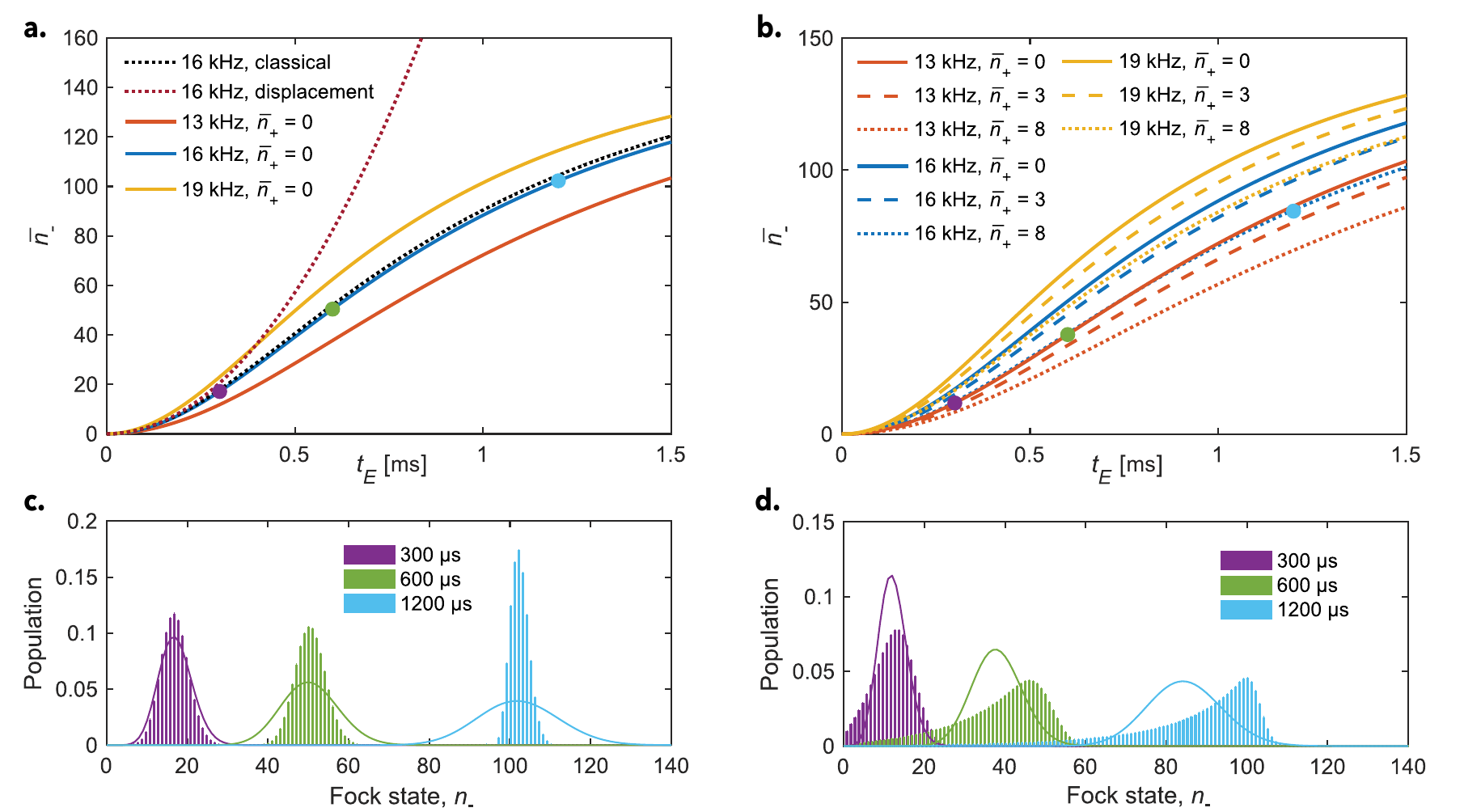}
    \caption{\textbf{Simulations of motional excitation.} Simulated average motional quantum numbers $\bar{n}_-$ of the ax-IP mode at different lattice interaction times for different ac-Stark shifts. \textbf{a.} No population in the spectator mode, \textbf{b.} with different thermal populations in the spectator mode. In the simulations, the detuning of the travelling lattice with respect to the target ax-IP frequency was set to zero, i.e. $\delta_- = 0$. The dotted lines in (a) correspond to excitation dynamics modelled with the classical model (dotted black trace) and employing a displacement operator (dotted red trace). See text for details. \textbf{c.} Phonon distributions of the target mode after applying an ODF corresponding to an ac-Stark shift of 16 kHz for different interaction times and no spectator mode population, and \textbf{d.} with a thermal population of $\bar{n}_+ = $ 8 phonons obtained after Doppler cooling in the ax-OP spectator mode. The times and colours match the coloured circles in (a) and (b). The solid lines indicate normalised Poissonian distributions expected if the motional excitations produced purely coherent motional states, as for the displacement model and as implied in the current classical treatment.}
    \label{fig:figure3}
\end{figure}

In Fig. \ref{fig:figure3} (a), the simulations for no spectator-mode population are shown. It can be seen that upon applying the ODF, the average phonon number of the target mode, $\bar{n}_-$, initially increases quadratically and then more slowly. In the early stage of the excitation when the ion is still in the Lamb-Dicke regime, $\left(\eta_{-}^{(1)}\right)^2 (2n_{-}+1) \ll 1$, the excitation is equivalent to applying a displacement operator $\hat{D}(\alpha) = e^{\alpha \hat{a}^\dagger - \alpha^* \hat{a} }$ with the coherent-state time-evolution defined as $\alpha= \eta_-^{(1)} E_{ac}^{0,1}t_E/\hbar$, which is shown as a dotted red trace. Such a displacement model is a first-order approximation to the Hamiltonian from Eq. (\ref{eq:quantum_hamiltonian1}) and is often used to describe coherent motional excitation \cite{leibfried03a, leibfried03b}. However, while this treatment describes the early stages of the time evolution satisfactorily, when the motional state populations follow a Poissonian distribution, it is not sufficient to describe the excitation outside of the Lamb-Dicke regime. Then, it is important to consider higher-order terms in the expansion of the Hamiltonian, which are included in our simulations. Under the experimental conditions presented here, this becomes relevant already for ODF pulse times $>$ 200 $\mu$s and $\bar{n}_-\gtrsim $ 15 phonons. As the average motional quantum number increases, so does the spread of the wave packet. As a result, the motional excitation is reduced, and the motional-state distribution becomes narrower compared to a coherent state with the same average phonon population, as shown in Fig. \ref{fig:figure3} (c). Such narrowing of the phonon distribution for highly excited motional states is expected to yield high-contrast Rabi flops. Similar motional excitation outside the Lamb-Dicke regime was previously employed to create squeezed states of motion \cite{mcdonnell07a, poschinger10a}.

It is instructive to compare the differences between the classical and quantum simulations. Such a comparison with no spectator mode population for the same ac-Stark shift of 16 kHz is shown by the dotted black and solid blue traces in Fig. \ref{fig:figure3} (a). Both treatments yield similar results for the initial dynamics, and the outcomes only slightly diverge with increasing interaction time. This finding indicates the small role of quantum effects for the considered excitation times. The classical model correctly recognises the behaviour outside of the Lamb-Dicke regime, i.e. reduction of motional excitation, when the particle's motion becomes comparable to the lattice wavelength. The agreement suggests that for fully suppressed populations in the spectator modes, a classical treatment would be sufficient.

From Figs. \ref{fig:figure3} (a) and (b), it is apparent that the excitation rate depends on the ac-Stark shift, which is directly proportional to the resultant ODF \cite{meir19a}. To detect smaller ac-Stark shifts, such as those originating from higher rotational states of N$_2^+$, it is required to increase the time of interaction with the lattice significantly to achieve comparable average final phonon populations. 

By contrast, modelling the effect of population in spectator modes necessitates the use of a quantum treatment. The presence of the spectator-mode population affects the motional excitation through the Debye-Waller effect which reduces the effective motional excitation of the ions in the target mode, as shown in Fig. \ref{fig:figure3} (b). The motional-state distribution of the target mode then significantly deviates from a Poissonian, especially at higher motional excitations, as can be seen in Fig. \ref{fig:figure3} (d) for an ax-OP spectator mode population of $\bar{n}_+$ = 8 phonons. Averaging over such a broad distribution of motional states significantly reduces the contrast of the resultant sideband Rabi flop that is important for reliable state detection. 

Note that since there is no cross-coupling between the target and spectator modes, and the modulation frequency of the travelling lattice is far detuned from the spectator modes, there is no increase in the spectator mode population. Consequently, in subsequent simulations of the experimental data, the spectator modes were assumed to exhibit constant thermal populations.

\begin{figure}[h] 
    \centering \includegraphics[width=1.0\textwidth]{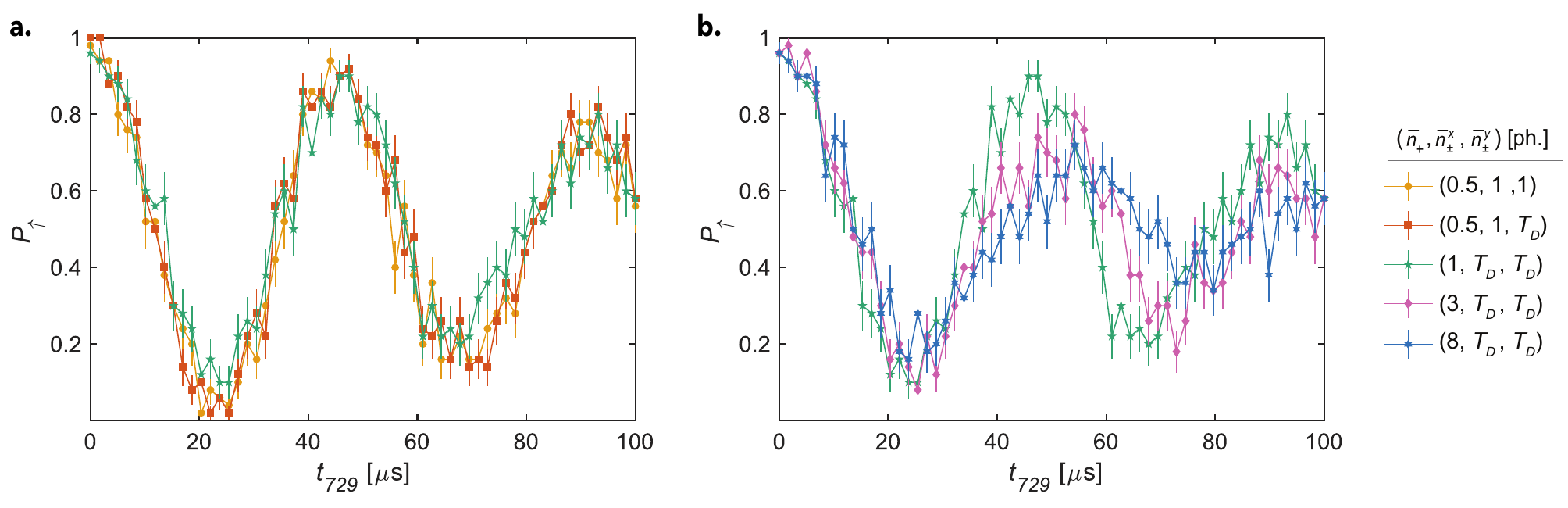} 
    \caption{\textbf{Rabi thermometry after motional excitation for different spectator mode populations.} Rabi oscillations on the blue sideband of the ax-IP mode on the $D_{5/2}(m_j=-5/2) \rightarrow S_{1/2}(m_j=-1/2)$ transition in Ca$^+$ after applying an ODF pulse for $t_E$ = 500 $\mu$s. The ax-IP target motional mode was initially cooled to $\sim 0.15$ phonons in all experiments. Different cooling methods were used to prepare defined average state populations in the ax-OP spectator mode, $\bar{n}_+$, and the radial modes $\bar{n}_\pm^x, \bar{n}_\pm^y$. $T_D$ indicates thermal populations after Doppler cooling of a motional mode. The green trace for ($\bar{n}_+$, $\bar{n}_{\pm}^x$, $\bar{n}_{\pm}^y$) = $(1,T_D,T_D)$ is the same in both plots. The effect of different temperatures in the radial modes is shown in \textbf{a.}, while the influence of the ax-OP mode population is shown in \textbf{b.}. Uncertainties represent the standard error of the mean for 50 experimental repetitions.} 
    \label{fig:rabi_flops_different_cooling}
\end{figure}

To make contact with experiment, Rabi oscillations measured on the blue sideband of the $D_{5/2}(m_j=-5/2) \rightarrow S_{1/2}(m_j=-1/2)$ transition in Ca$^+$ transition following an ODF pulse used to probe the internal state of the molecule are shown for different spectator-mode populations in Fig. \ref{fig:rabi_flops_different_cooling}. First, consider the effect of different populations in the radial modes while both axial modes were initially cooled close to their ground states (Fig. \ref{fig:rabi_flops_different_cooling} (a)). The minute change in Rabi frequency between the green and yellow/red traces can be attributed to the small but non-negligible difference in the ax-OP spectator mode population. In addition, the small loss of contrast around $t_{729}$ = 24 $\mu$s may be related to differences in the final motional-state-population distribution between the cooling methods employed in the experiments (EIT cooling -- yellow/red traces, SB cooling -- green trace data). Besides these small differences, it is evident that any effect of radial-mode populations is negligible at the sensitivity limit of the present experiments: no major difference was observed whether both, only a single or none of the radial modes were cooled close to the ground-state. This result is expected because the coupling (described by the Lamb-Dicke parameters) between the lattice laser beams propagating along the axial direction and the radial motional modes is negligible. If any influences of radial-mode populations would have been detected, they would have indicated other effects such as couplings between the axial and radial modes caused by, e.g., the Kerr-like effects or trap asymmetries.  

By contrast, the effects of population in the ax-OP spectator mode on the excitation-dynamics of the ax-IP mode were found to be significant. We controlled the spectator-mode population by varying the number of pulses in the SB cooling sequence. The effects of $\bar{n}_+ \sim$ 8 phonons (blue trace), 3 phonons (pink trace) and 1 phonon (green trace) on the Rabi flops on the blue ax-IP sideband are compared in Fig. \ref{fig:rabi_flops_different_cooling} (b). The differences in the Rabi flops cannot be solely attributed to the DW factors from Eq. (\ref{eq:DW_plus}) which modify the effective Rabi frequencies (Eq. (\ref{eq:rabi_freq_thermometry})) in the spectroscopy. These DW factors directly contribute to the sideband readout signal, and were estimated to decrease the effective Rabi frequency by only a few percent \cite{sinhal20a}. 

\begin{figure}[t]
    \centering \includegraphics[width=1\textwidth]{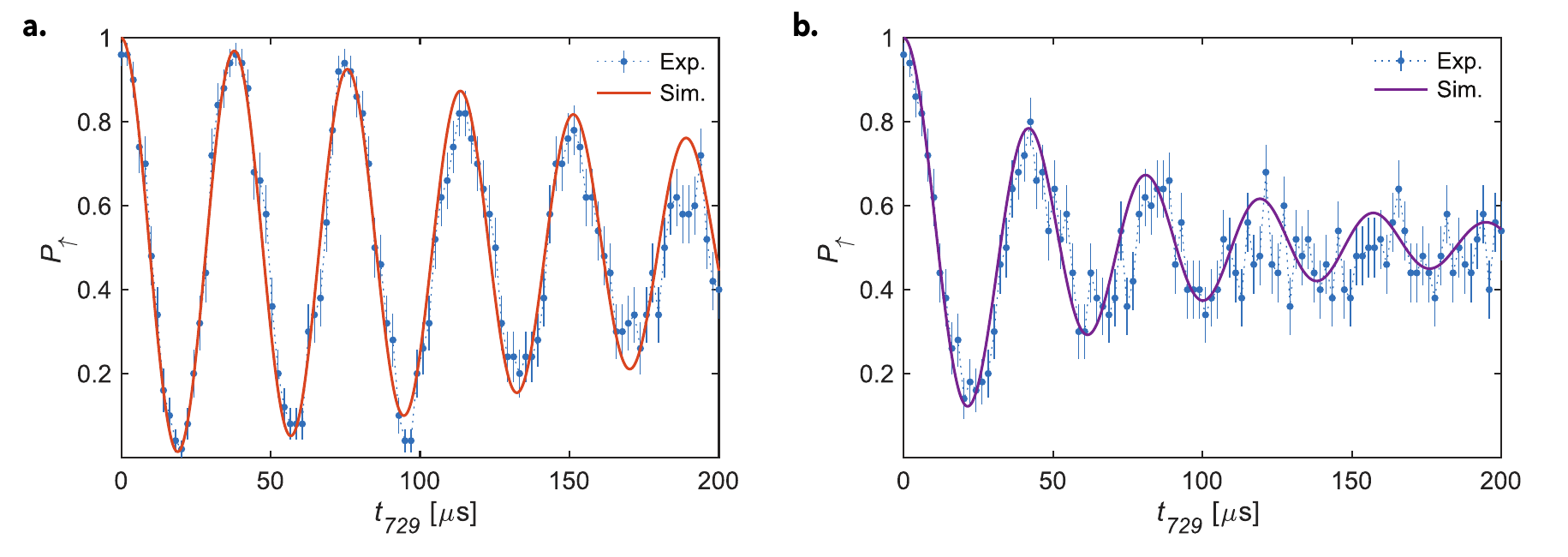} 
    \caption{\textbf{Comparison of simulations with experiment.} Rabi oscillations on the blue ax-IP sideband of the $D_{5/2}(m_j=-5/2) \rightarrow S_{1/2}(m_j=-1/2)$ transition in Ca$^+$ after applying a state-dependent ODF on N$_2^+$ for $t_E$ = 500 $\mu$s. The measurements correspond to ax-OP spectator mode populations of \textbf{a.} $<$ 0.5 ph., and \textbf{b.} $\sim 8$ ph. The solid lines correspond to theoretical Rabi flops computed with Eq. (\ref{eq:probability_excited_state_decoherence}) and the motional-mode population distributions $P_{n_+,n_-}$ extracted from the simulations. Uncertainties represent the standard error of the mean. See text for details.}
    \label{fig:rabi_flops_fit_data}
\end{figure}

The results of the Rabi thermometry strongly depend on the populations of the axial modes, $P_{n_+,n_-}$, as evidenced in Eq. (\ref{eq:rabi_oscillations_eq1}). We carried out quantum simulations of the lattice interacting with N$_2^+$ at different ac-Stark shifts and at the relevant experimental parameters (lattice time, temperature of the modes) from which $P_{n_+,n_-}$ were computed. The resulting theoretical Rabi flops were compared with the experimental data. In Fig. \ref{fig:rabi_flops_fit_data}, experimental Rabi flops (a) after initially cooling both axial modes close to their ground states and (b) no secondary cooling of the ax-OP mode following Doppler cooling (SB$_{\text{no}\, \text{ax-OP}}$ in Tab.~\ref{tab:frequencies_and_temperatures}) are compared with simulations. By comparing both experiments, it can be seen that excitations in the ax-OP spectator mode markedly reduce the contrast of the Rabi oscillations, which is well reproduced by the simulations. The reason can be traced back to the DW factors Eqs.~(\ref{eq:DW1}) and (\ref{eq:DW2}) which are effective in the excitation by the optical lattice described by Eq.~(\ref{eq:quantum_analytical_sol}). 

The contrast of the resulting Rabi flop, and therefore the state-detection fidelity, depends on the shape of the generated motional wave packet. Fig. \ref{fig:figure2} (c) shows the wave packets corresponding to the flops in Figs. \ref{fig:figure2} (a) and \ref{fig:rabi_flops_fit_data} (a). The dependence of the Rabi frequencies on the motional quantum number $n_-$ of the target mode is plotted in Fig. \ref{fig:figure2} (b). The high-contrast flop in Fig. \ref{fig:figure2} (a) originates from a narrow wave packet with an average phonon number $\bar{n}_-=88$ in a region where the Rabi frequency is only weakly dependent on $n_-$. Thus, the Rabi frequencies for excitations out of all $n_-$ states populated in the wave packet are similar which, by virtue of Eq. \ref{eq:rabi_oscillations_eq1}, results in a Rabi flop that nearly resembles an excitation out of a single state. Conversely, the flop in Fig. \ref{fig:rabi_flops_fit_data} (a) corresponds to a broader wave packet in Fock space averaging at $\bar{n}_-=39$ where the dependence of the Rabi frequency on $n_-$ is more pronounced, leading to loss of contrast in the Rabi flop. 

The exact ac-Stark shift in the simulation was extracted from the best match of the simulations with the experiment when both axial modes were cooled to their ground states. The ac-Stark shift of $\Delta E^0_{ac}= (16 \pm 1)$~kHz on N$_2^+$ (from a single lattice beam) used in the simulations is close to the (17.5 $\pm$ 1.0)~kHz calculated for the lattice-beam intensities (2.14 $\pm$ 0.12 W/mm$^2$) estimated in the experiments \cite{sinhal21thesis}.  

For the purpose of molecular-state identification, it is sufficient to probe the Rabi flop around the $\pi$-time of the blue sideband where the highest signal-to-background ratio can be achieved (interval indicated by the black dotted lines in Fig.~\ref{fig:figure2} (a)) \cite{sinhal20a}. A main motivation for improving cooling and understanding the excitation dynamics is to increase the contrast between signal and background (see, e.g., Fig. \ref{fig:figure2} (a)) and therefore improve the state-detection fidelities. The advantages are twofold. First, better cooling of the axial modes reduces the background signal. Second, the amplitude of the Rabi oscillations when cooling normal modes close to their ground states was noticeably increased due to suppression of DW effects. The resulting state detection fidelity can be calculated using the experimental data from Fig. \ref{fig:figure2} (a) by following the treatment outlined in Refs. \cite{meir19a} and \cite{sinhal20a}. For a Rabi $\pi$-time of $t_{\pi}$=15.3~$\mu$s, the probability of detecting population transfer on the blue sideband with no lattice excitation (either background signal or N$_2^+$ in excited state) was $P_{\downarrow}^{bg.}$=0.05, while with the lattice excitation, when N$_2^+$ was in the ground state, was $P_{\downarrow}^{sig.}$=0.95. Under these conditions, from the statistical analysis, we compute a state-detection fidelity exceeding 99.99\% for as few as nine experimental repetitions. Note, however, that the actual state-detection fidelity achieved in the experiment may be lower because of systematic effects such as laser drifts and collisions with background gas (see also Ref. \cite{chaffee25a}). Thus, depending on the targeted fidelity, the number of measurements and therefore the duration of the experiment can be significantly reduced, e.g., comparing to Ref.~\cite{sinhal20a}, the experimental cycle improved by a factor of two. This is particularly important for room-temperature experiments with (molecular) species whose chemical lifetime is limited by reactions with background gases. Shorter exposure to the lattice lasers is also desirable due to the potential for off-resonant scattering of the ground molecular state. Both of these factors play a role in the experiments presented here. 

\begin{figure}[htbp]
    \centering
    \includegraphics[width=0.6\textwidth]{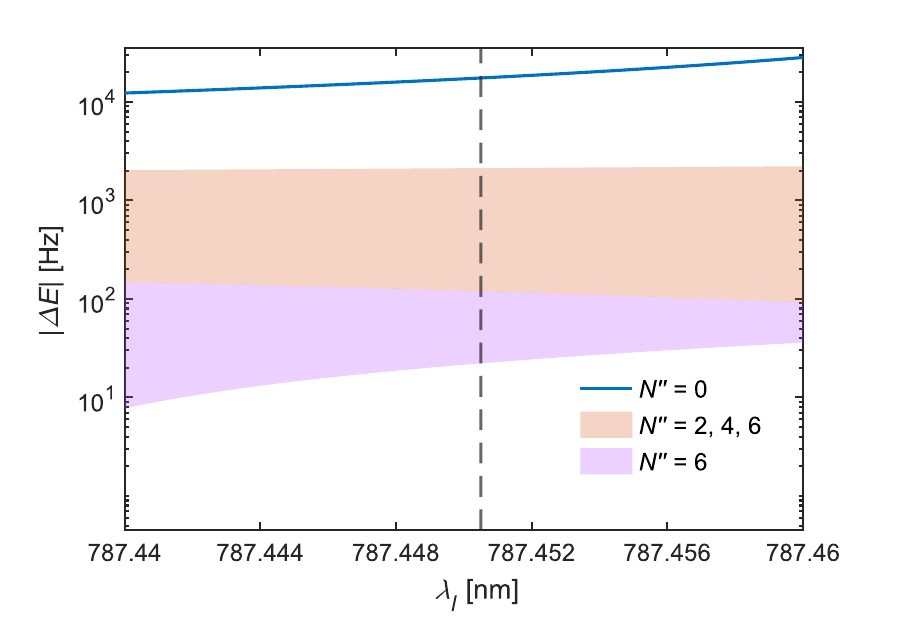}
    \caption{\textbf{ac-Stark shifts of molecular states at different optical-lattice wavelengths.} Calculated absolute magnitude of ac-Stark shifts for $ A^2\Pi_u(v'=2, J', F', MF') \leftarrow X^2\Sigma_g^+(v''=0, N'', J'', F'', MF'')$ transitions from the four lowest rotational states $N^{\prime\prime}$ of ortho-N$_2^+$ ($I=0, 2$) assuming a single lattice laser with an intensity of 2.14~W/mm$^2$ at different wavelengths, $\lambda_{l}$, of relevance for the present experiments. The black dashed line indicates the wavelength of 787.4505~nm employed for the optical lattice in the present experiments. See text for details.}
    \label{fig:shifts_from_rot_transitions}
\end{figure}

The increase in sensitivity thus achieved allows for detection of much smaller ac-Stark shifts which could originate from the population of rotationally excited states of N$_2^+$. In Fig.~\ref{fig:shifts_from_rot_transitions}, the calculated ac-Stark shifts are shown for different rotational states of N$_2^+$ and their individual hyperfine manifolds \cite{sinhal21thesis, najafian20a, najafian20b}. Note that the REMPI scheme used to produce N$_2^+$ is isomer-selective and produces only ions with nuclear spins quantum numbers $I=0$ and 2, and thus yields only even rotational states \cite{shlykov23a}. Hyperfine transitions from the rotational ground state, $N^{\prime\prime}$=0, for the laser frequency used in this experiment, experience similar ac-Stark shifts (within $<$ 0.5 kHz) and are indicated by the blue trace. Almost all hyperfine transitions from $N^{\prime\prime}$=2, 4 exhibit ac-Stark shifts of a similar magnitude in the wavelength range shown and are indicated by the red-shaded area in Fig. \ref{fig:shifts_from_rot_transitions}. The ac-Stark shifts of hyperfine transitions from  $N^{\prime\prime}$= 6 lie within the red- and violet-shaded areas. The ac-Stark shifts for the states $N^{\prime\prime}$ $>$ 0 are at least one order of magnitude weaker compared to $N^{\prime\prime}$=0 at the lattice wavelength indicated by the vertical black dashed line. Higher excited rotational states with $N^{\prime\prime}$$>$6 are unlikely to be produced with the employed REMPI scheme. As demonstrated in Ref. \cite{najafian20b}, the range of possible states can be narrowed down by comparing the measured ac-Stark shifts with theory and taking advantage of the phase sensitivity of the effective ODF. In principle, the sensitivity for the detection of molecular states corresponding to smaller ac-Stark shifts can be further enhanced by increasing the lattice interaction time without the need to adjust other parameters of the optical lattice, such as it's wavelength, intensity or polarization. 

\begin{figure}[t]
    \centering 
    \includegraphics[width=1.0\textwidth]{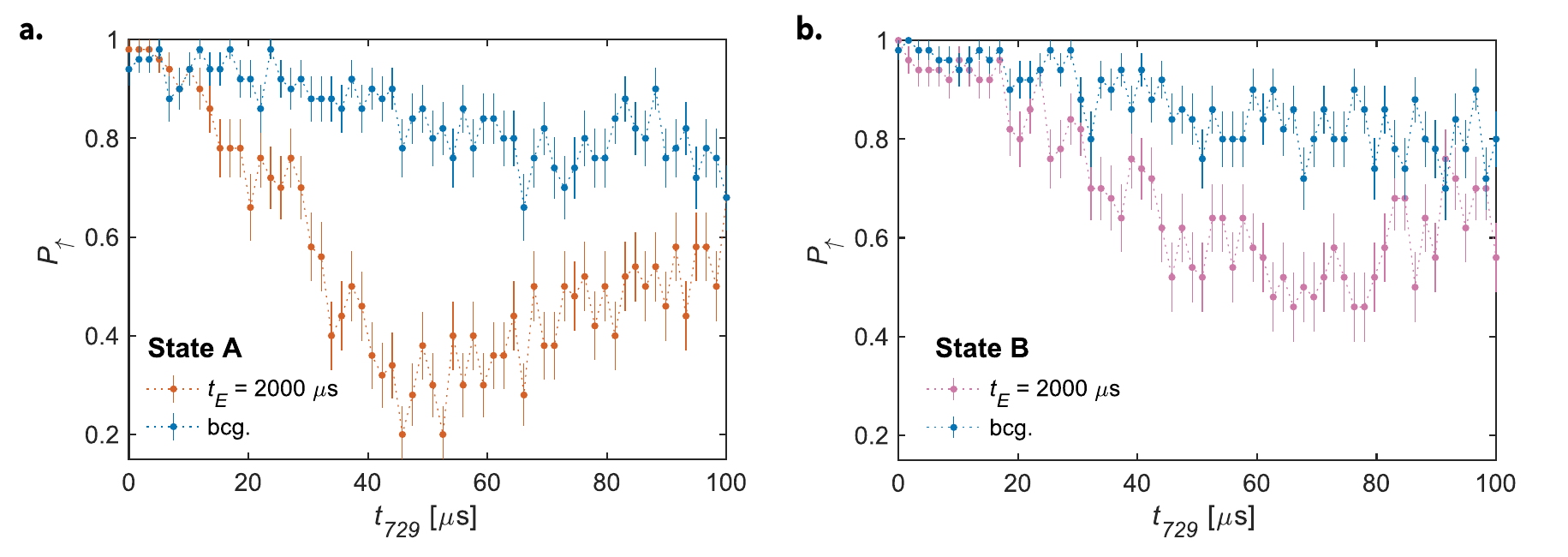}     
    \caption{\textbf{Detection of excited rotational states.} Rabi oscillations on the blue sideband of the $D_{5/2}(m_j=-5/2) \rightarrow S_{1/2}(m_j=-1/2)$ transition in Ca$^+$ after applying the optical lattice on N$_2^+$ in two different rotational states (\textbf{a.} and \textbf{b.}) corresponding to the $N''=2, 4, 6$ manifold. Lattice excitation time was $t_E=2000\,\mu s$ (orange and pink). The background traces (blue) were obtained without applying the lattice fields. Uncertainties represent the standard error of the mean.}
    \label{fig:higher_rotational_states}
\end{figure}

In Fig. \ref{fig:higher_rotational_states}, we report experiments which correspond to such a detection of excited rotational states. The ac-Stark shifts extracted from these measurements are compatible with the orange-shaded area displayed in Fig. \ref{fig:shifts_from_rot_transitions}. We emphasise that the improved signal-to-background ratio after cooling all normal modes and an increased lattice-ion interaction time (here $t_E = $ 2000 $\mu s$) was a prerequisite for detecting the weak ODF produced by these higher rotational states at the present lattice parameters.

\section{Conclusions}\label{sec:conclusions}
In conclusion, it was shown that spectator modes play an important role in the dynamics of motional excitation of a two-ion string in a trap with an optical lattice used for the quantum-state detection of molecular ions. Supported by simulations, we showed that this can be attributed to Debye-Waller effects in the interaction of the ions with the lattice. By cooling the spectator modes of the two-ion string close to their ground states, we improved the signal-to-background ratio in molecular-state detection and achieved a statistical fidelity exceeding 99.99\% within only nine experimental repetitions. As a result, compared to our previous report \cite{sinhal20a}, the total experimental time was reduced by a factor of two, in addition to the enhanced detection fidelity. This reduction is particularly valuable when probing short-lived molecular states and for minimizing off-resonant scattering from the lattice lasers. Furthermore, the improved sensitivity of the current scheme opens up new possibilities for the identification of excited rotational states at lattice parameters optimised for the detection of the ground state.

This work provides insights into the behaviour of trapped ions interacting with an optical dipole force. It underlines the importance of cooling spectator modes in similar quantum-non-demolition state-detection experiments, as well as in other experiments involving manipulating quantum states encoded in motional degrees of freedom, such as bosonic quantum computing with trapped ions. The marked improvement in readout fidelity demonstrated here has a direct impact and benefit for all advanced quantum protocols necessitating state detection as well as related applications such as molecular spectroscopy and frequency metrology \cite{najafian20b}.

\bmhead{Acknowledgements}
We thank Mudit Sinhal and Gregor Hegi for their contributions to the development of the theory employed in this paper. We also thank the referee for his insights that resulted in the addition of Figure \ref{fig:figure2} (b) and (c). The present work was supported by the Swiss National Science Foundation (grant nr. 200021\_204123), the Eurpoean Partnership on Metrology (Funder ID: 10.13039/100019599, grant nr. 23FUN04 COMOMET), and the University of Basel.

\bmhead{Author contributions}
M.R. carried out the experiments and simulations. A.S. contributed to the development and characterization of the experimental setup. Z.M. developed the theory behind the simulations. S.W. conceived and supervised the project. M.R and S.W. drafted the manuscript, all authors contributed to the final version.

\bmhead{Competing interests}
The authors declare no competing interests.

\bmhead{Data availability}
The primary data underlying the findings of the present study are available on Zenodo at https://zenodo.org/records/15234656.

\bmhead{Code availability}
The software codes used to generate the simulations in the present study are available on Zenodo at https://zenodo.org/records/15234656.

\backmatter

%% \bmhead{Supplementary information}
%% If your article has accompanying supplementary file/s please state so here.

%% APPENDICES
\begin{appendices}

\section{Derivation of Eq. (\ref{eq:quantum_analytical_sol})}\label{sec:appendix1}

In this section, we outline the derivation of Eq. (\ref{eq:quantum_analytical_sol}) from Eq. (\ref{eq:quantum_hamiltonian1}). Details on the description of normal modes of a Coulomb crystal of two ions with unequal masses in a harmonic trap can be found in Refs. \cite{morigi01a, home11a}. 

Consider the 1D model of a travelling optical lattice interacting with a two-ion string described by the Hamiltonian $\hat{H}$ from Eq. (\ref{eq:quantum_hamiltonian1}). The motion of the ions can be described in terms of two normal modes --  in-phase (denoted with `$-$') and out-of-phase (denoted with `$+$') mode, with the frequencies given in Eq. (\ref{eq:mode_freq_eq1}).

First, the Hamiltonian from Eq. (\ref{eq:quantum_hamiltonian1}) is separated into time-independent and -dependent parts:
\begin{equation}
    \hat{H} = \hat{H}_0 + \hat{H}_I (t) ,
\end{equation}
where
\begin{eqnarray}
    \hat{H}_0=  \hbar\omega_{-}(\hat{a}^\dagger_{-}\hat{a}_{-}+ \frac{1}{2}) +\hbar\omega_{+}(\hat{a}^\dagger_{+}\hat{a}_{+} + \frac{1}{2}) ,\\
    \hat{H}_I(t)= \sum_{j=1,2} 2\Delta E_{ac}^{0,j}(1+\cos(2k\hat{z}_j-\omega_l t)) .
\end{eqnarray}

The position operators $\hat{z}_j$ for ions $j={1,2}$ are related to normal mode coordinates $\hat{z}_\pm$ by:
\begin{equation}
\label{eq:position_operators}
\begin{aligned}
& \hat{z}_1=\sqrt{\mu}\left(\hat{z}_{+} \cos \theta+\hat{z}_{-} \sin \theta\right)+z_1^{init.}, \\
& \hat{z}_2=-\hat{z}_{+} \sin \theta+\hat{z}_{-} \cos \theta+z_2^{init.},
\end{aligned}
\end{equation}
with $z_j^{init.}$ as the equilibrium position of the ions and the angle $\theta$ defined as in Eq. (\ref{eq:tan_theta_eq}).

In the interaction picture, the normal-mode position operators become:
\begin{equation}
\begin{aligned}
\label{eq:interaction_position_operators}
& \hat{z}_{+} \rightarrow \hat{z}_+^{\prime}(t)=z_{+}^{0}\left(a_{+}^{\dagger} e^{i \omega_+ t} + a_{+} e^{-i \omega_+  t}\right) ,\\
& \hat{z}_{-} \rightarrow \hat{z}_{-}^{\prime}(t)=z_{-}^{0}\left(a_{-}^{\dagger} e^{i \omega_-t}+a_{-} e^{-i \omega_-t}\right) ,
\end{aligned}
\end{equation}
where $z_{\pm}^{0}$ was defined in Eq. (\ref{eq:spatial_extent_wf}).

The interaction Hamiltonian in the interaction picture becomes \cite{najafian20b}:
\begin{eqnarray}
\label{eq:interaction_hamiltonian}
\hat{H}_I^{\prime}(t) &&= \sum_{j=1,2} 2 \Delta E_{ac}^{0,j} \left( 1+ \cos \left(2 k \hat{z}_j^{\prime}  -\omega_l t \right)\right) \nonumber\\
&&\approx \sum_{j=1,2} 2 \Delta E_{ac}^{0,j} \left(\cos \left(2 k \hat{z}_j^{\prime}  -\omega_l t    \right)\right) \nonumber\\
&&= \sum_{j=1,2} \Delta E_{ac}^{0,j} \left(\exp\left(i(2k \hat{z}_j^{\prime}  - \omega_l t )\right) + c.c. \right) . 
\end{eqnarray}
Here, the constant term in the first line is an energy shift that does not contribute to the dynamics and was thus omitted without loss of generality. 

The dynamics of the two-ion crystal interacting with the modulated optical-dipole force is given by the Schr\"odinger equation:
\begin{equation}
\label{eq:schrodinger_eq}
    i \hbar \partial_t |\psi_I(t)\rangle =\hat{H}_I^{\prime}(t) |\psi_I(t)\rangle \,.
\end{equation}
The motional wavefunction is defined on the 2D Fock space of two normal modes:
\begin{equation}
\label{eq:wavefunction_states}
     |\psi_I(t)\rangle = \sum_{n_+, n_-} C_{n_+, n_-}(t) |n_+, n_- \rangle
\end{equation}
with the time-dependent coefficients $ C_{n_+, n_-}(t)$. 

By combining Eq. (\ref{eq:schrodinger_eq}) with Eqs. (\ref{eq:interaction_hamiltonian}), (\ref{eq:position_operators}), (\ref{eq:interaction_position_operators}) and (\ref{eq:wavefunction_states}) one obtains:
\begin{eqnarray}
    i \hbar \dot{C}_{m_+, m_-}(t) &=& \sum_{n_+, n_-}C_{n_+, n_-} \langle m_+, m_- |  
    \left[ \Delta E_{ac}^{0,1} \left(\exp\left(2 i k (\sqrt{\mu}\left(\hat{z}_{+}^{\prime} \cos \theta+\hat{z}_{-}^{\prime} \sin \theta\right)))  - i\omega_l t + i\phi_1  \right) + c.c. \right) \right]  \nonumber\\
    &&+ \left[\Delta E_{ac}^{0,2} \left(\exp\left(2 i k (\left(-\hat{z}_{+}^{\prime} \sin \theta+\hat{z}_{-}^{\prime} \cos \theta \right)))  - i\omega_l t + i\phi_2  \right) + c.c. \right) \right] |n_+, n_-\rangle \,.
    \label{eq:appendix_state_coefficients_1}
\end{eqnarray}
Here, the initial position of the ions was incorporated into the phase shifts $\phi_j$.
\\
\\
To simplify the treatment, three assumptions are made:
\begin{enumerate}
    \item The target mode can only change by $\pm$ 1 phonon at a time. 
    \item The spectator mode population is not affected by the ODF pulse. 
    \item There is no phonon exchange between the modes.
\end{enumerate}
Points 1 and 2 are valid for small detunings of the lattice-modulation frequency from the target-mode frequency, when it is much smaller than the mode frequencies, the lattice frequency and the detuning from the spectator mode, i.e. $|\delta_-| \ll \omega_+, \omega_-, \omega_l, |\delta_+|$. 
\\
\\
To simplify the Eq. (\ref{eq:appendix_state_coefficients_1}), we use the relation \cite{wineland79a, leibfried03a}:
\begin{equation}
    \label{eq:motional_contributions_exp}
    \langle n+s |  e^{i\,\eta(a+a^{\dagger})} |n\rangle = e^{-\eta^{2}/2}\eta^{\left|s\right|} \sqrt{\frac{n_{<}!}{n_{>}!}}L_{n_{<}}^{\left|s\right|}(\eta^{2}).
\end{equation} 
where $s$ is the change of the number of phonons and $n_{<(>)}$ is the lesser (greater) of $n$ and $n+s$. From assumptions 1 and 2 above, $s=\pm1$ for the $n_-$ mode and $s=0$ for the $n_+$ mode. Other transitions between motional states are neglected. 
\\
In the rotating-wave approximation, we thus get:
\begin{eqnarray}
i\hbar \dot{C}_{n_+,n_-}(t) &=& \sum_{j=1,2} \Delta E_{ac}^{0,j} e^{-(\eta_+^{(j)})^2/2} L_{n_+}^0((\eta_+^{(j)})^2) e^{-(\eta_-^{(j)})^2/2} \eta_-^{(j)}  \nonumber \\
&& \times \Bigg[C_{n_+,n_- -1}e^{-i\delta_- t+i\phi_j} \frac{L_{n_- -1}^1((\eta_-^{(j)})^2)}{\sqrt{n_-}} \nonumber \\
&&+\, C_{n_+,n_- +1}e^{+i\delta_- t-i\phi_j} \frac{L_{n_-}^1((\eta_-^{(j)})^2)}{\sqrt{n_- +1}} \Bigg] ,
\end{eqnarray}
with Lamb-Dicke parameters defined according to Eqs. (\ref{eq:new_def_LD_params1}-\ref{eq:new_def_LD_params4}).
By rearranging this equation and using the definitions of DW factors from Eq. (\ref{eq:DW_plus}), we finally obtain Eq.~(\ref{eq:quantum_analytical_sol}).
\end{appendices}

%% BIBLIOGRAPHY
\bibliography{BibFile}

\end{document}